# Particle Acceleration by Electromagnetic-Dominated Outflows

*Edison Liang and Koichi Noguchi, Rice University, Houston TX 77005-1892*

## ABSTRACT

We review recent developments in particle acceleration by Poynting flux using plasma kinetic simulations, and discuss their potential applications to gamma-ray burst phenomenology.

## 1. INTRODUCTION

An outstanding problem in modern astrophysics is the acceleration of high-energy particles The challenge is to find natural, robust mechanisms which efficiently convert electromagnetic, rotation, hydrodynamic, thermal or gravitational energy, into the relativistic kinetic energy of a small number of nonthermal particles. The potential applications of particle acceleration include cosmic rays, high-energy radiation from pulsars, blazars, gamma-ray bursters (GRBs), magnetars/SGRs and black holes (BHs).

It has long been speculated that many high-energy astrophysical phenomena (e.g. pulsar winds, GRBs) may be the result of particle acceleration and radiation by electromagnetic (EM)-dominated outflows ("Poynting flux" (Lyutikov and Blandford 2003, Lyutikov and Blackman 2003, Smolsky and Usov 2000)). However, until recently there had been few concrete models of such acceleration as they require large-scale Particle-in-Cell (PIC, (Birdsall and Langdon 1991)) simulations of relativistic collisionless plasmas. Three years ago we embarked on a pioneering study of Poynting flux acceleration (PFA) using 2.5D (2D-space,3-momenta) PIC codes originally developed at LANL (Nishimura et al 2002) and LLNL (Langdon and Lasinski 1976). This led to the discovery of a novel phenomenon called diamagnetic relativistic pulse accelerator (DRPA, (Liang et al 2003), which represents the first explicit demonstration of efficient, robust PFA from first principles plasma kinetics. Since then we have made rapid advances along this path, including the discovery of a fundamental scaling law for the Lorentz factor, long term evolution of the distribution function and pulse profile, radiation output and damping. A remarkable result is the striking similarity between the asymptotic properties of DRPA and observed signatures of GRBs (Liang and Nishimura 2004), including their light curves, spectra and spectral evolution (Fishman and Meegan 1995, Preece et al 2000). Recently we also demonstrated the robustness of DRPA in 3D, and that PFAs with generic initial conditions lead to similar asymptotic states as DRPA (Liang 2005).

## 2. DRPA AS EXAMPLE OF PFA

When a long vacuum EM wave with wavelength $\lambda \gg c/\omega_{pe}$ ($\omega_{pe}$= electron plasma frequency) irradiates a plasma surface, it induces a diamagnetic skin current $\mathbf{J}$ that shields the EM field from entering the plasma and reflects the incident wave. The resultant $\mathbf{J} \times \mathbf{B}$ force, called ponderomotive force, accelerates surface particles into the plasma, away from the EM pulse (Fig.1a). For an electron-positron (e+e-) plasma and large EM amplitude ($\Omega_e/\omega_{pe} \gg 1$, $\Omega_e$=eB/mc=electron gyrofrequency), the ponderomotive force accelerates the upstream plasma to relativistic energies (Wilks et al 1992), and the reflection front "snowplows" the entire plasma to a maximum Lorentz factor $\gamma_{max} \sim (\Omega_e/\omega_{pe})^2$ via energy-momentum conservation (Kruer et al 1975). Conversely, when a long EM pulse with $\lambda \gg c/\omega_{pe}$ imbedded inside a plasma tries to escape, the diamagnetic skin current $\mathbf{J}$ inhibits the EM field from leaving. In this case the $\mathbf{J} \times \mathbf{B}$ force captures and accelerates the surface plasma into the EM pulse (Fig.1b). As the EM pulse "pulls" the plasma behind it, it is slowed by the plasma loading (group velocity < c). In contrast to the snowplow case above, here the EM pulse continually sheds slower particles that fall behind. Consequently, the EM pulse, together with the fastest particles, accelerate with time as its plasma loading decreases. We call this phenomenon diamagnetic relativistic pulse accelerator (DRPA).

For a slab e+e- plasma, >60% of EM energy is eventually converted into the directed energy of the fastest particles (Fig.2a), and the asymptotic $\gamma_{max} >> (\Omega_e/\omega_{pe})^2$ (Liang et al 2003, Liang and Nishimura 2004). For an e-ion plasma however, charge separation caused by ion inertia transfers most of the EM energy into ion energy, and few electrons are accelerated (Fig.2b). DRPA is a nonlinear, collective relativistic phenomenon, which has no analog in the weak field ($\Omega_e/\omega_{pe} < 1$), low density ($\lambda < c/\omega_{pe}$) or test particle limit.

## 3. ASYMPTOTIC PROPERTIES OF DRPA

Fig.3 highlights the global evolution of a pure e+e- EM-dominated expansion into a vacuum in both slab and cylindrical geometries, showing the trapping and acceleration of the surface particles at late times by the EM pulse. At late times the DRPA exhibits a number of remarkable properties: repeated bifurcation of the density profile, development of a power-law momentum distribution with low-energy cut-off, and growth of the peak Lorentz factor with the square-root of the number of gyroperiods. These results have important implications for astrophysics, especially GRBs (Sec.4). Fig.4 illustrates the detailed pulse structure evolution. As the density pulse advances, it bifurcates repeatedly due to nonlinear coupling between longitudinal and transverse EM modes. In these simulations the bifurcation process begins at t ~10$L_o$/c ($L_o$= initial plasma width), leading to a complex multi-peak structure at late times. We have analyzed the Fourier spectra of late-time density profiles. Preliminary results suggest that the highest Fourier peaks have wavelengths ~ $c\gamma_m(t_b)$/(few $\omega_{pe}$), where $\gamma_m(t_b)$ is the Lorentz factor of the EM pulse at the time of bifurcation $t_b$. The bifurcated pulses strongly resemble GRB light curves (see Sec.5) (Fishman and Meegan 1995).

As the pulse of Fig.4 advances, a peak develops in the momentum distribution of the pulse particles (Fig.5). This peak Lorentz factor $\gamma_m = p_{xmax}/m_e c$ corresponds to the group velocity of the EM pulse (< c due to plasma loading). Particles whose x-momenta lie below $p_{xmax}$ gradually lose acceleration and dephase. This creates the deficit of low-energy particles in the pulse momentum distribution function (Fig.5 and Fig.4 phase plots). At $\Omega_e t$ >5000 a power law develops above $p_{xmax}$. In this example the power-law slope of -3.5 is close to the particle momentum index of many astrophysical -ray sources. The most important result emerging from these long-duration simulations is the growth of the peak Lorentz factor $\gamma_m(t)$ with t according to:

$$\gamma_m(t) = (2f\Omega_e(t)t + C_o)^{1/2} , \qquad t > L_o/c \qquad (1)$$

where $C_o$ and f are constants dependent on initial conditions (Fig.6, (Liang and Nishimura 2004)). Eq.(1) can be derived analytically using Landau-Lifshitz (Landau and Lifshitz 1965).

## 4. DRPA AND GRB PHENOMENOLOGY

Though DRPA represents only a special idealized example of PFA, it reproduces four unique signatures of GRBs: (a) complex and extremely diverse light curves (Fishman and Meegan 1995), (b) ubiquitous power-law spectra with a spectral break (Preece et al 2000), (c) hard-to-soft spectral evolution in single-peak FRED bursts (Fishman and Meegan 1995); (d) correlation of spectral hardness with intensity in complex, multi-peak bursts

Fig.4 provides a plausible explanation for the diversity of GRB light curves, if we translate the electron density profile into γ-ray light curves via ray-tracing. If a DRPA radiates all of its energy early in the expansion (Fig.4a, e.g. strong B or running into dense stellar photons), it produces a smooth single-peaked GRB (i.e. a "Fast Rise Exponential Decay" pulse or FRED (Fishman and Meegan 1995). If it radiates late, after repeated bifurcation (Fig.4d-e), it produces a GRB with complex, multiple peaks. Of course, multiple peaks may also involve multiple events from the central engine, or interactions with CSM/ISM (Sec.6.2). The momentum distribution in Fig.5 is consistent with typical GRB spectra, since particle index=3.5 translates into photon index=2.25 (Preece et al 2000), and the spectral break is due to deficit in low-energy particles in the front pulse. Fig.4 also explains the overall hard-to-soft spectral evolution. Fig.4a shows that for unbifurcated smooth pulses, spectral hardness peaks before the photon flux peak, in agreement with observations (Preece et al 2000). Fig.4d,e show that narrow peaks in complex pulses tend to be more symmetric than FREDs (Norris et al 1996, 2000), and hardness and intensity are correlated in complex, multi-peak pulses (Norris et al 1996, 2000).

The critical frequency of intrinsic DPRA radiation can be estimated using (Landau and Lifshitz 1965): $\omega_{cr} \sim \gamma_m \Omega_e$. Eq.(1) plus $\omega_{cr} \sim \gamma_m \Omega_e$ lead to a simple scaling law for spectral break energy $E_{pk} \sim B^{3/2}$. If the volume of the emitting shell varies little from burst to burst, then we obtain the scaling law $E_{pk} \sim U_{em}^{3/4}$ where $U_{em}$ is total EM energy injected into the burst. Data from GRBs of known z indeed show that $E_{pk} \sim U_{\gamma}^{3/4}$ where $U_{\gamma}$ is total $\gamma$-ray output corrected for beam angle (Ghirlanda et al 2004), so this correlation is worth further study using Swift data. Eq.(1) also leads to meaningful numbers. Suppose that typical GRB radiation time $t_r \sim 300s$, since most GRBs have already bifurcated (Fishman and Meegan 1995). Eq.(1) plus $h\omega_{cr}(t_r)/2\pi \sim 500$ keV for $E_{pk}$ (Preece et al 2000) give $\gamma_m \sim$ few.$10^7$ and $B \sim 10^6$G for $f \sim O(1)$. $B \sim 10^6$G then gives $U_{em}(t_r) \sim 10^{50}$ ergs, assuming a $4\pi$ shell of thickness $L(t_r) \sim 10^{12}$ cm and radius $R \sim ct_r \sim 10^{13}$ cm, and total initial energy $U_{tot} \sim 10 U_{em}(t_r) \sim 10^{51}$ ergs (cf.Fig.2a), consistent with GRB energetics (Piran 2000, Meszaros 2002). With the above values for B and $\gamma_m$, we find for the widths of bifurcated subpulses in long complex GRBs $\sim \gamma_m/(few.\omega_{ep}) \sim$ ms to 10's of ms, also consistent with observations (Fishman and Meegan 1995).

## 5. ROBUSTNESS OF PFA

In the original simulations of DRPA (Liang et al 2003, Liang and Nishimura 2004) (Figs.2,3) the initial condition was modeled as two equal counter-propagating linearly polarized EM pulses with aligned **B** vectors (so that **E**=0=**J** at t=0) imbedded in a uniform plasma. This clever representation of Poynting fluxes as unconfined static B fields without supporting currents was simple to implement in PIC codes, but it is somewhat artificial and difficult to realize in nature. Recently we have simulated PFAs with more realistic initial conditions. Two examples are (a) confined **B** field with supporting currents **J**=curl **B** at t=0; (b) opposing vacuum EM pulses that penetrate a central plasma by first compressing it to a thickness < 2 skin depths. In both cases we find that the asymptotic state of accelerated plasma is similar to DRPA (Figs.7,8) (Liang 2005). This confirms that the asymptotic properties of DRPA are representative of more generic PFAs.

Another concern was whether 3D effects may inhibit or destabilize the DRPA mechanism. This concern was dismissed with new 3D simulations (Liang 2005, Noguchi et al 2005). Fig.9 shows a sample 3D (10x10x500) simulation of slab e+e- expansion with the same input parameters as in Figs.2,3. We see no evidence of any instability in the y-z plane, and the 3D phase plots and spatial profiles evolve identically to the 2.5D results. Physically, this is because the driver of DRPA acceleration at late times is the x-component Lorentz force $v_z$ x $B_y$. The momentum $p_y$ conjugate to the symmetry coordinate y plays no role in the acceleration. Hence relaxing the symmetry condition in y does not affect the evolution of the accelerator.

## 6. Radiation from the accelerated particles

For astrophysical applications we need to know how accelerated particles radiate. PIC codes are useful for computing radiation output since they accurately track the velocity and acceleration history of every particle. But standard PIC codes only include wave-particle interactions in the mean-field approximation They do not treat the loss or scattering of high-energy radiation with $\lambda <<$ Debye length. Recently we implemented self-consistent radiation and radiation damping (RD) in the NN code by adding the Dirac radiation reaction force $f_D$ (Noguchi et al 2005). We have compared the radiation power output using $(-v.f_D)$ to the relativistic dipole formula (Rybicki and Lightman 1979) and found excellent agreements. We therefore replace the Lorentz equation in the NN code with the Dirac-Lorentz equation, using Landau-Lifshitz approximation (Landau and Lifshitz 1965). This allows for self-consistent calculation of particle acceleration with RD in the classical and optically thin limit. This approximation is valid as long as $B < 4x10^{13}$ G and self-absorption is negligible. The latter condition is easy to satisfy for PFA because both the radiation intensity and the accelerated particle momenta are highly anisotropic with $p_x >> p_z >> p_y$.

Figs.10,11 compare the evolution of the DRPA (same initial conditions as Fig.3a) with and without RD. In the RD case EM energy is transferred to radiation and the particles act as transfer agents with little change in energy. In this run the field is set high ($r_e\Omega_e/c=10^{-3}$, $r_e$=classical electron radius) so that we can see the effect of RD within $t\Omega_e<10^4$. Even for such high fields, the largest RD effect is at the beginning when transverse Lorentz force and synchrotron radiation dominate. At late times when the longitudinal force dominates, RD is small, and a parasitic calculation using the non-RD PIC code gives acceptable results (Fig.11). Fig.10 shows that density bifurcation is enhanced by RD and radiation damps both $p_x$ and $p_z$.

Using ray-tracing to account for light-path differences from different parts of the source, we have developed a post-processing routine to compute the intensity and polarization measured by detectors at infinity as functions of view angle (Noguchi et al 2005). Fig.12 gives sample results of such output. As expected the detected radiation is strongly linearly polarized since the driving Poynting flux is linearly polarized. The small depolarization comes from the initial y-momentum dispersion which is small since the initial plasma is cold. For an initially hot plasma the polarization will be weaker. In the coming year, we plan to apply this to runs with finite source sizes in the y-z plane to visualize how the source evolution (Fig.4a) translates into detected light curves (Fig.4b) as functions of the view angle. The next project is to compute the detected radiation spectrum using FFT of individual particle velocity history (Jackson 1975), before summing all particles with ray tracing.

The above classical treatment of radiation in PIC codes does not include Compton scattering of external photons, which ultimately must be computed with Monte Carlo (MC) methods using the quantum Klein-Nishina cross-section (Boettcher and Liang 2001). Sugiyama (private communications) has recently implemented an approximate treatment of inverse Compton loss in the NN code by adding a time-averaged damping force $f_D= -4\sigma_T\gamma^2 v U_s/3c$, where $\sigma_T$= Thomson cross-section and $U_s$=soft photon energy density, to represent the net effect of scattering by many isotropic soft photons (Rybicki and Lightman 1979). Fig.13 gives sample results of this code.

## 7. INTERACTION WITH AMBIENT PLASMA

Astrophysically it is important to study the interaction of PFAs with the ambient environment, both to see how the ambient plasmas damp the acceleration and absorb the Poynting flux, and how the heated ambient plasma radiate. During this past year we have performed a number of e+e- DRPA runs into cold ambient e+e- and e-ion plasmas. Fig.14 shows sample results of 2.5D runs. Some of our major findings include: (a) the maximum Lorentz factors achieved even with ambient plasma are comparable to the vacuum case of Fig.4, though the number of high-energy particles decreases with increasing ambient density. (b) a complex multi-phase plasma exists in the contact region where high-γ ejecta plasma overruns low-γ "swept-up" plasma. The swept-up plasma consists of 2 phases: a cold unaccelerated phase coexisting with a PF-accelerated phase whose Lorentz factors < those of the ejecta. In the e-ion case, both the "swept-up" electrons and ions exist in 2 distinct phases. The transition region is very broad (>>> the ion gyroradii), contrary to MHD results. (c) the swept-up ion Lorentz factors are << the electron Lorentz factors because ions are pulled only by the charge separation electric fields, while leptons are accelerated by the EM pulse directly. This disproves the conventional MHD assumption that the electron and ion bulk Lorentz factors are the same and the transition layer thickness is ~ ion gyroradius. (d) there is no evidence of any disruptive plasma instability (Weibel, 2-stream, and lower-hybrid-drift instabilities) at the contact interface. Such instabilities are suppressed to first order by the strong transverse EM field. (e) the Poynting flux decays via ponderomotive acceleration of the ejecta and ambient electrons, plus mode conversion to longitudinal plasma waves which are absorbed by Landau damping.

## 8. PF ACCELERATION OF IONS

In a pure e-ion plasmas most of the EM energy goes into the ions due to charge separation (Fig.2b) (Nishimura et al 2003). Since ions are massive they are poor $\gamma$-radiators. The situation is different when we have a mixture of e-ion and e+e- plasmas. The EM pulse selects out the mobile e+e- component and preferentially accelerates them, leaving the e-ion component behind (Fig.15). There is little charge separation in the e-ion component since the e+e- component is neutral. Consequently almost all of the EM energy goes into the e+e- component and the ions gain little energy. This solves one of the major problems faced by other acceleration mechanisms since any realistic e+e- plasma will likely be contaminated with some ions.

The above result suggests that if PFA is operating in GRBs and other $\gamma$-ray sources, then a bright $\gamma$-ray source is likely e+e- dominated and not a prolific source of HECRs. Conversely, a prolific HECR source should be deficient in e+e- and therefore not a bright $\gamma$-ray source. Such distinction may be testable with future coordinated GLAST and ground-based observations. The mixture ratio of e-ion to e+e- needed to effectively couple the two components likely depends on the input parameters, such as $\Omega_e/\omega_{pe}$, kT and $L_o$.

## Figure Captions

**Fig.1** *(left) Schematics showing the **J x B** forces (black arrow) when an EM pulse tries to enter a plasma (a) and when it tries to leave a plasma (b). DRPA is based on the second scenario*

**Fig.2** *Time evolution of the magnetic field, electric field, and particle energies for (a) the electron-positron DRPA expansion; (b) the electron-ion expansion (from Liang et al 2003).*

**Fig3** *(left) 2.5D PIC simulations of slab and cylindrical $e+e-$ plasma acceleration, with initial plasma temperature $kT=5$ MeV, $\Omega_e/\omega_{pe}=10$ ($\Omega_e=eB_o/m_e$, $\omega_{pe}=(ne^2/\varepsilon_o m_e)^{1/2}$), initial slab width $L_o=120c/\Omega_e$ and uniform internal **B**$=(0, B_o, 0)$. We show the $x\geq0$ snapshots of particle distribution (a-d), axial magnetic field (color scale runs from $B_y=+0.2B_o$ (red) to $-0.1B_o$(blue)) and current density (white arrows) for the cylindrical case (e), and phase plot for the slab case (f). $\Omega_e t=800$ for all left panels and $\Omega_e t=10^4$ for all right panels. The green dot in the phase plot denotes the initial phase volume (from Liang and Nishimura 2004)*

**Fig.4** *(right) Particle density profiles (blue curves, right scales) and phase plots (red dots, left scales) for the slab run of Fig.1 at (a)$\Omega_e t=1000$, (b)5000, (c)10000 and (d)18000. Current densities are plotted in the small insets. Note the repeated bifurcation of the density peak. Panel (e) is the $\Omega_e t=30000$ snapshot of another run with $L_o=600c/\Omega_e$, showing many more peaks. These results should be compared to the GRB light curves of Fig.4b. The "hard-to-soft" trend of the momentum distribution is clearest for panel (a): the momentum peaks before the density, in agreement with BATSE data for FREDs (from Liang and Nishimura 2004).*

**Fig.5** *(l) Evolution of the x-momentum distribution for all surface particles in the slab DRPA of Fig3a, showing the peak Lorentz factor $\gamma_m(=p_{xmax}/m_ec)$ increasing with time. and a power law of slope $\sim -3.5$ at $\Omega_e t=6000$ (from Liang and Nishimura 2004).*

**Fig.6** *(r) The peak Lorentz factor $\gamma_m$ versus time for the slab pulse, compared with Eq.(1). The best-fit curve (dotted) gives $f=1.33$ and $C_o=27.9$. Note that $\Omega_o(t).t =3800$ is equivalent to $\Omega_e.t=18000$ due to B decay. This result shows that for $t>L_o/c$, $\gamma_m$ scales as the square-root of the number of gyroperiods (from Liang and Nishimura 2004).*

**Fig.7 (l)** *Evolution of $B_y$, $J_z$ and phase plot of $e+e-$ slab expansion as in Fig.3a, but with initial **J=curl B**. The late time phase plot is almost identical to that of DRPA (here $t=t\Omega e/30$) (Liang 2005).*

**Fig.8 (r)** *Late-time phase plot of $e+e-$ slab accelerated by colliding vacuum EM pulses (right) is almost identical to that of DRPA (left) (Liang 2005)*

**Fig.9** *3D (10x10x500) DRPA simulation with the same input parameters as in Fig.3a. Both the particle distribution (blue dots) and field amplitudes (color contours) remain uniform in the y-z plane (Noguchi et al 2005)*

**Fig.10** *(left) Phase plots and density profiles of a no-RD run compared to an RD run. (same input parameters as in Fig.3a (from Liang 2005).*

**Fig.11** *(right) Comparison of average radiation output per electron computed from the runs of Fig.10 (from Noguchi et al 2005).*

**Fig.12** *Detected radiation intensities and polarization as functions of detector time for $\Omega_e/\omega_{pe}$ =10 and $\Omega_e/\omega_{pe}$ =100 runs including RD using ray-tracing (upper panels). Bottom panel is detected angular fluence plotted against view angle ($\theta$ in x-z plane and $\phi$ in x-y plane) (Noguchi et al 2005).*

**Fig.13** *Evolution of $\Omega_e/\omega_{pe}$ =100 DRPA phase plots including Comptonization of external blackbody soft photons. Left panel is for low photon density and right panel is for high photon density, showing the rapid decline of the maximum Lorentz factors (Sugiyama unpublished).*

**Fig.14** *Time-lapse phase plots and B-field profiles for a slab e+e- 2.5D DRPA running into (0.001x ejecta density) cold e-ion ambient plasma. $P_i$ is in units of $m_i c$. The acceleration of ejecta plasma stalls after it has swept up roughly equal mass of ambient plasma but the acceleration of ambient ions continues due to the pull by charge separation Ex. EM field decays by acceleration of swept-up plasma and mode conversion. Lower right panels show the blow-up details of the transition region (from Liang 2005).*

**Fig.15** *Phase plots of the electrons and ions for a mixture of 10% e-ion and 90% e+e- plasma compared to pure e-ion plasma (from Liang 2005).*

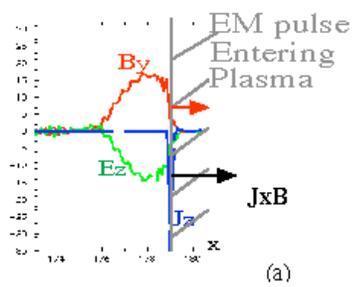

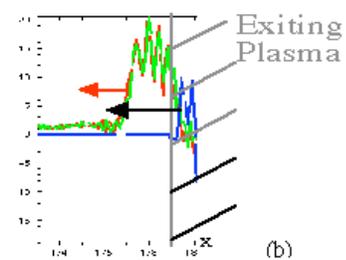

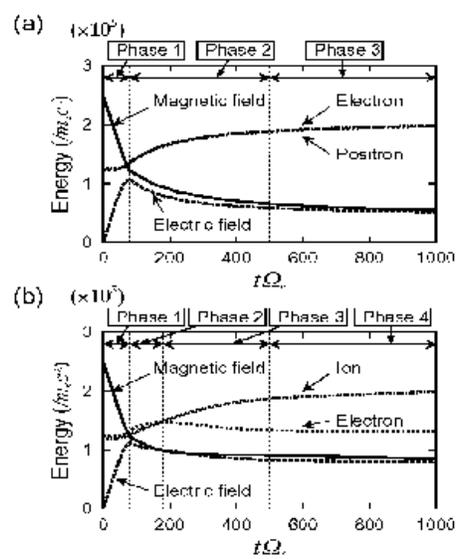

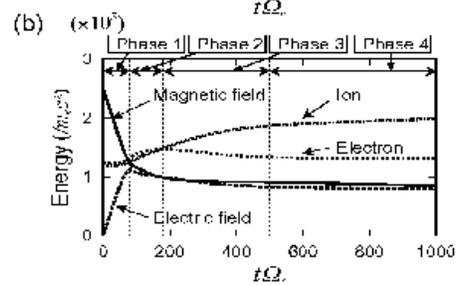

Fig.1 (left) , Fig.2 (right)

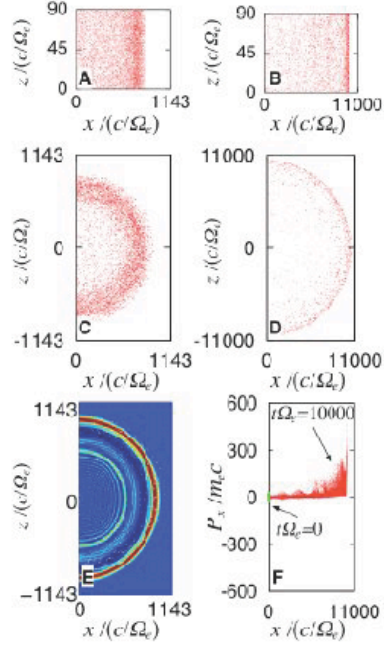

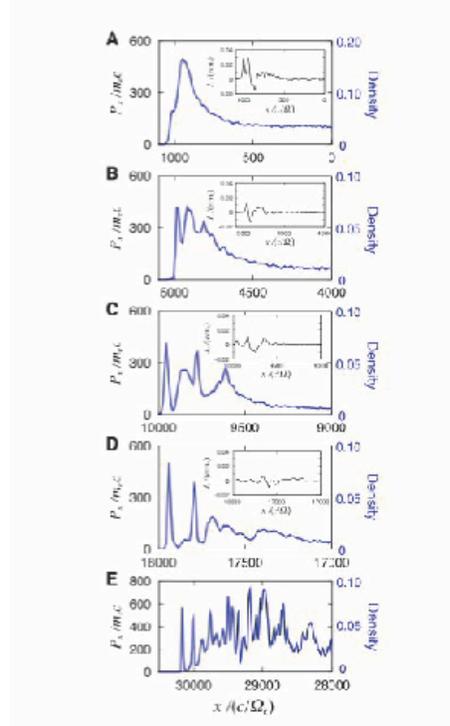

Fig3 (left) ,Fig.4 (right)

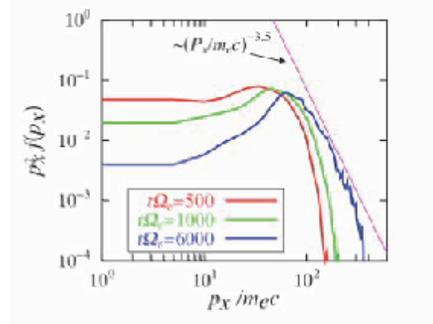

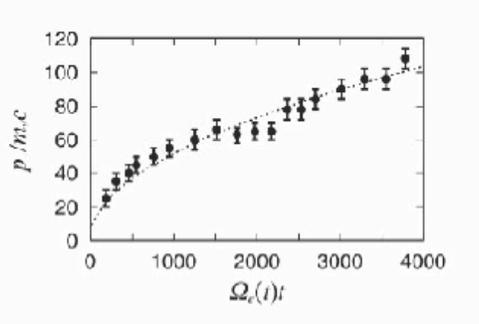

Fig.5 (l), Fig.6 (r)

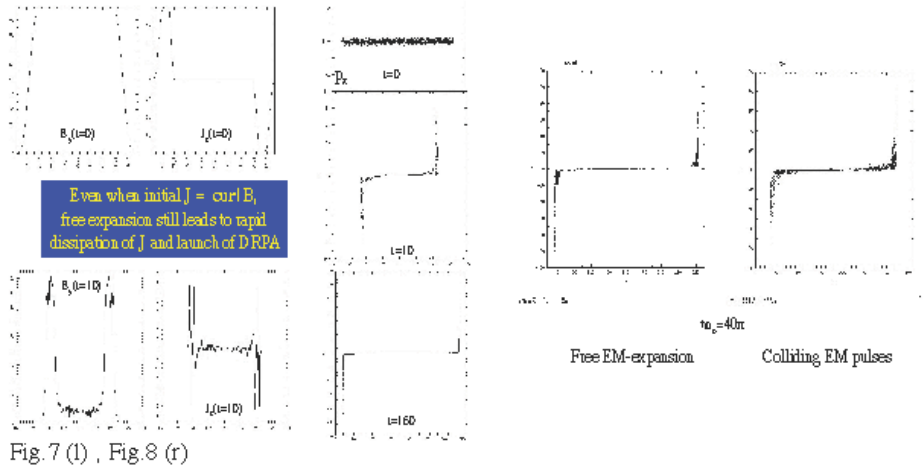

Even when initial $J = \text{curl } B$, free expansion still leads to rapid dissipation of $J$ and launch of DRPA

Fig.7 (l) , Fig.8 (r)

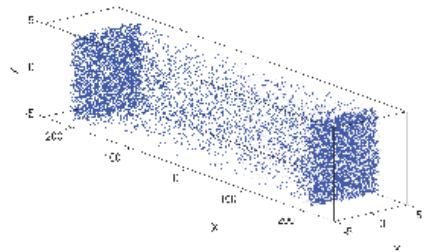

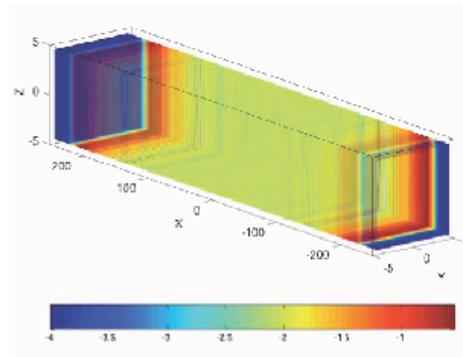

Free EM-expansion          Colliding EM pulses

Fig.9

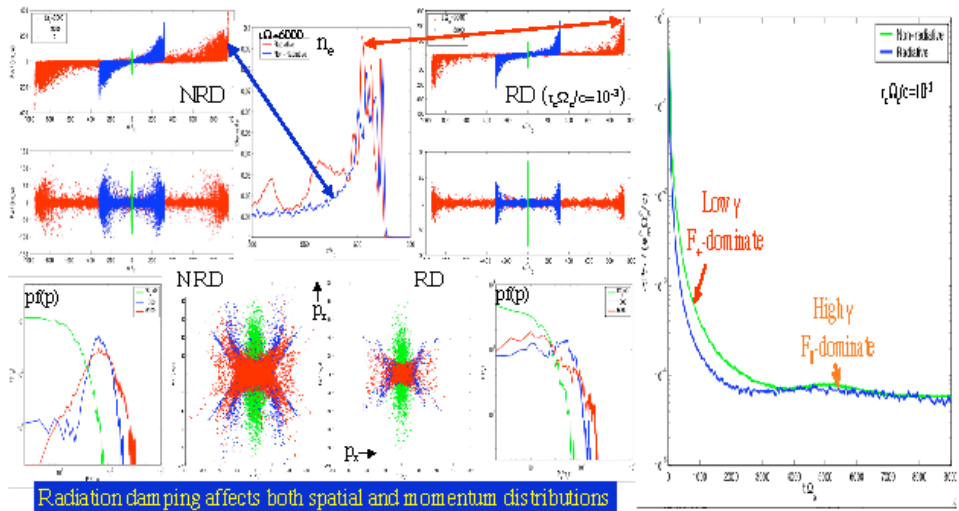

Radiation damping affects both spatial and momentum distributions

Fig.10 (left) ,Fig.11 (right)

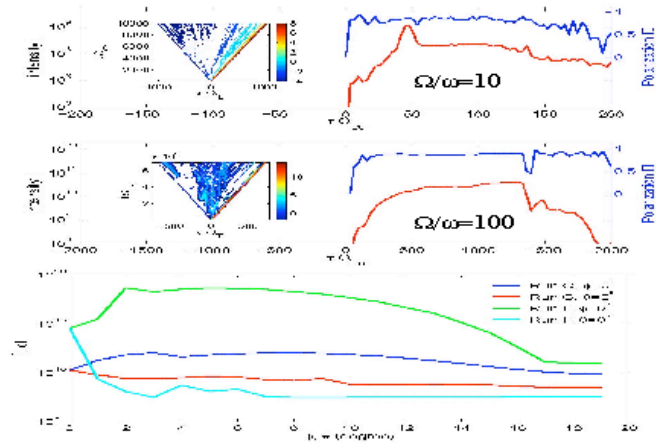

Fig.12

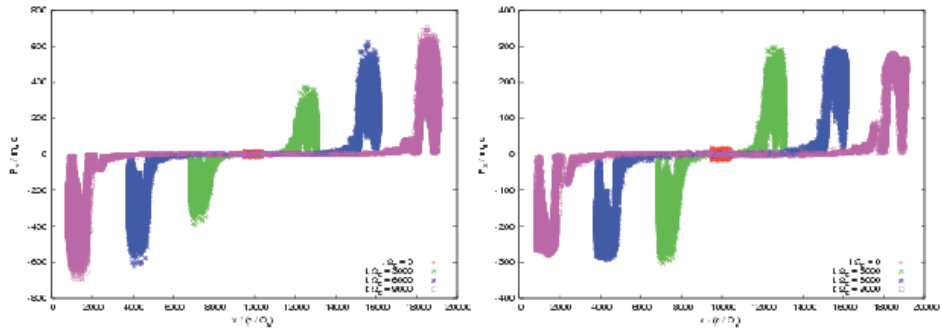

Fig.13

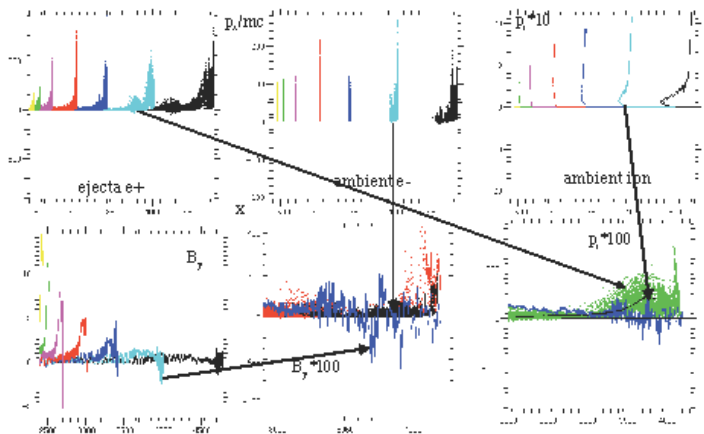

Fig.14

In mixture of e-ion and e+e- plasma, EM pulse selectively accelerates only the e+e- component

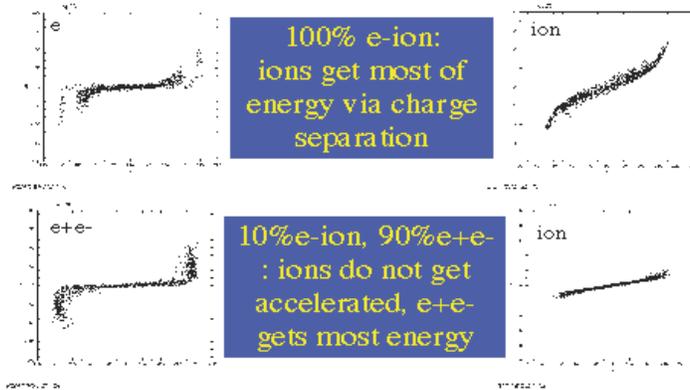

100% e-ion: ions get most of energy via charge separation

10%e-ion, 90%e+e-: ions do not get accelerated, e+e- gets most energy

Fig.15